\documentclass[12pt]{article}

\usepackage{graphicx}

\def\beq   {\begin{equation}}
\def\eeq   {\end{equation}}
\def\bea {\begin{eqnarray}}
\def\eea {\end{eqnarray}}

\def\psla{p\kern-.45em/}

\def\sq{\ifmmode{\tilde{q}} \else{$\tilde{q}$} \fi}
\def\sg{\ifmmode{\tilde{g}} \else{$\tilde{g}$} \fi}

\def\dr{\ifmmode{\overline{\rm DR}} \else{$\overline{\rm DR}$} \fi}
\def\ms{\ifmmode{\overline{\rm MS}} \else{$\overline{\rm MS}$} \fi}
\def\drbarr{\ifmmode{\overline{\rm DR}'} \else{$\overline{\rm DR}'$} \fi}

\begin{document}


\vspace*{-1cm} 
\begin{flushright}
  TU-746 \\
  hep-ph/0506262
\end{flushright}

\vspace*{1.4cm}

\begin{center}

{\Large {\bf
Two-loop SUSY QCD correction to the \\
gluino pole mass
} 
}\\

\vspace{10mm}

{\large Youichi Yamada}

\vspace{6mm}
\begin{tabular}{l}
{\it Department of Physics, Tohoku University, Sendai 980-8578, Japan}
\end{tabular}

\end{center}

\vspace{3cm}

\begin{abstract}
\baselineskip=15pt
The pole mass of the gluino is diagrammatically calculated to the 
two-loop order in SUSY QCD as a function of the running 
parameters in the lagrangian, for the gluino and squarks 
sufficiently heavier than the quarks. 
The $O(\alpha_s^2)$ correction shifts the gluino mass 
by typically 1--2 \%, 
which may be larger than the expected accuracy 
of the mass determination at future colliders. 
\end{abstract}

\vspace{20mm}

\newpage
\pagestyle{plain}
\baselineskip=15pt

If supersymmetry (SUSY) with the breaking scale around 
1 TeV is a solution to the hierarchy problem between the 
electroweak scale and the Planck/grand unification scale, 
all particles in the standard model have their 
superpartners with masses not much higher than the electroweak scale. 
These new particles will then be produced 
at colliders in near future, such as the CERN Large Hadron Collider (LHC) 
and 
the International Linear Collider (ILC). 
Precision studies of these particles will be possible 
at these colliders \cite{lhcilc}. 

One of the main motivations for these studies is the 
investigation of the SUSY breaking mechanism in the unified theory. 
The soft SUSY breaking parameters may be extracted from precision 
measurements at future colliders \cite{msusydeterm,SPA}, and 
extrapolated to much higher scale \cite{softextrapo,softextrap}
by renormalization group equations. 
Structure of the soft SUSY breaking at higher scale provides 
an important clue to the SUSY breaking mechanism in the unified theory. 
For example, 
the unification of three gaugino masses at the same scale as 
that of the gauge couplings 
is crucial for the SUSY grand unified theory \cite{susygut1,susygut2} and 
superstring phenomenology \cite{string}. 

For this purpose, we need not only precise measurements of 
physical parameters such as masses and cross-sections, 
but also precise prediction of the relations between 
these observables and parameters in the lagrangian. 
Sometimes we have to calculate the relations beyond the one-loop 
order to match expected precision at future experiments. 
To the masses of the particles, for example, 
two-loop corrections to the top and bottom quarks \cite{mq2loopsusy}, 
squarks in the first two generations \cite{msq2loop}, 
and Higgs bosons \cite{mh2loop} have been calculated 
in the framework of the minimal SUSY standard model (MSSM) \cite{mssm}. 

Here we consider the correction to the mass of 
the gluino $\tilde{g}$, the superpartner of the gluon, in the MSSM. 
At the CERN LHC, gluino is, if it is sufficiently light, 
expected to be copiously produced by the strong interaction \cite{sgprod}. 
A study \cite{msgdet1} has shown that, 
for the SUSY parameter set SPS1a given in Ref.~\cite{SPS1a} 
with $m_{\sg}\simeq 600$ GeV, $m_{\sg}$ may be determined 
to accuracy $\delta m_{\tilde{g}}=8$ GeV from 
precision data at the LHC, and even to $\delta m_{\tilde{g}}=6.5$ GeV 
when combined with the data from the ILC. 
On the other hand, the one-loop QCD correction to the 
gluino mass \cite{mv1,pierce1,pierce2} is much larger, typically $O(10)$ \%, 
due to large coupling constant and large SU(3) representation of the gluino. 
One naively expect the two-loop correction to $m_{\sg}$ is $O(1)$ \%, 
similar order to the experimental uncertainty. 
It is therefore important to examine whether higher-order corrections 
to the gluino mass is relevant in extracting SUSY breaking terms 
at future precision experiments. 
In addition, these corrections might be relevant to the 
uncertainty \cite{uncertainty} in the calculation of the SUSY 
particle masses by the computer codes \cite{codes} for 
given values of the soft SUSY breaking terms at the unification scale, 
at the same order as other two-loop mass corrections and 
the three-loop contributions \cite{3loop} to the running of parameters 
between the unification scale and the scale of the SUSY particle masses. 

In this paper, we calculate the pole mass of the gluino 
as a function of the lagrangian parameters, 
including SUSY QCD correction to $O(\alpha_s^2)$, 
by diagrammatic calculation. 
For simplicity, we ignore quark masses and 
squark mixings in the loops. This approximation is valid 
for the case where the gluino and squarks are sufficiently 
heavier than the quarks. We also 
assume degenerate mass $m_{\sq}$ for squarks. 
The effects of the quark masses and left-right mixings of squarks 
to the gluino mass correction will be briefly commented later. 

The pole mass $m_{\sg}$ of the gluino, which is defined in terms of 
the complex pole $s_p=(m_{\sg}-i\Gamma_{\sg}/2)^2$ of 
the gluino propagator, is given at the two-loop order by 
\begin{eqnarray}
m_{\sg} 
&=& {\rm Re} \frac{ M_3 - \Sigma_M(s_p) }{ 1 + \Sigma_K(s_p) } 
\nonumber\\ 
&=& M_3 - {\rm Re}[ M_3\Sigma_K^{(1)}(M_3^2)+\Sigma_M^{(1)}(M_3^2) ]
- {\rm Re}[ M_3\Sigma_K^{(2)}(M_3^2)+\Sigma_M^{(2)}(M_3^2) ]
\nonumber \\ 
&& + {\rm Re} \left[ 
\{ M_3\Sigma_K^{(1)}(M_3^2)+\Sigma_M^{(1)}(M_3^2) \}
\{ \Sigma_K^{(1)}(M_3^2) + 2M_3^2\dot{\Sigma}_K^{(1)}(M_3^2) 
+ 2M_3\dot{\Sigma}_M^{(1)}(M_3^2) \} \right] 
\nonumber\\ 
&\equiv& M_3 + \delta m_{\sg}^{(1)} +\delta m_{\sg}^{(2)} . 
\label{pole} 
\end{eqnarray}
Here $M_3$ is the running tree-level gluino mass in the lagrangian. 
$\Sigma_{K,M}^{(1)}$ and $\Sigma_{K,M}^{(2)}$ are the one-loop 
and two-loop parts of the gluino self energy 
\begin{equation}
i\Sigma(p) = i(\Sigma_K(p^2)\psla + \Sigma_M(p^2) ) ,
\end{equation}
respectively. 
The dot in Eq.~(\ref{pole}) denotes the derivative with respect to 
the external momentum squared $p^2$. 

The SUSY QCD contribution to $\Sigma(p)$ is generated by loops with 
the gluino, gluon, quarks, and squarks. 
Parameters in the lagrangian 
are renormalized in the \drbarr scheme \cite{drbarprime}, which is 
the \dr scheme (dimensional reduction \cite{DR} with 
modified minimal subtraction) with additional finite 
counterterms for squark masses to 
remove the dependence on the mass $m_{\epsilon}$ of 
the $\epsilon$-scalar \cite{escalarmass}. 

The one-loop correction $\delta m_{\sg}^{(1)}$ 
in our approximation of massless quarks and degenerate 
squarks is \cite{pierce1,pierce2} 
\beq
\delta m_{\sg}^{(1)} = 
\frac{C_V \alpha_s }{4\pi} M_3 \left( 5 -6\log \frac{M_3}{Q} \right) 
+\frac{\alpha_s }{\pi} N_q T_F 
M_3 B_1(M_3^2, 0, m_{\sq}) ,
\label{oneloop}
\eeq
where $C_V=3$, $T_F=1/2$, and $N_q=6$ is the number of quarks. 
Parameters ($\alpha_s$, $M_3$, $m_{\sq_i}$) 
in Eq. (\ref{oneloop}) are the \drbarr running ones at the 
renormalization scale $Q$. 
$B_1(p^2,m_1,m_2)$ is the finite parts of 
the one-loop Passarino-Veltman function \cite{PV}, defined by 
($D=4-2\epsilon$) 
\beq
B_1(p^2, m_1, m_2) = 
\frac{1}{p^2} \frac{Q^{2\epsilon}}{\Gamma(1+\epsilon)} 
\int\frac{d^Dk}{i\pi^{D/2}} 
\frac{ k\cdot p }{(k^2-m_1^2) [(k+p)^2-m_2^2] } 
+\frac{1}{2\epsilon} . 
\eeq
Its explicit form for $m_1=0$ is 
\beq
B_1(M_3^2,0,m_{\sq}) = 
\frac{(m_{\sq}^2-M_3^2)^2}{2M_3^4}\log\frac{m_{\sq}^2-M_3^2}{m_{\sq}^2}
+\frac{1}{2}\log\frac{m_{\sq}^2}{Q^2}+\frac{m_{\sq}^2}{2M_3^2} -1. 
\eeq

\begin{figure}[ht]
\begin{center}
\includegraphics[width=11cm]{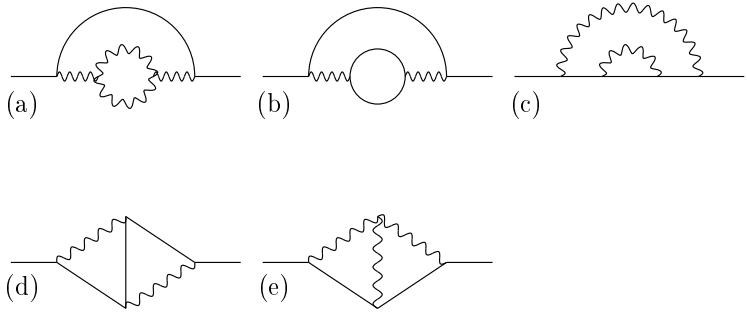}
\end{center}
\caption{ 
\small Two-loop $O(\alpha_s^2)$ contributions to the gluino self energy, 
without quark and squark propagators. 
}
\label{fig1}
\end{figure}

\begin{figure}[ht]
\begin{center}
\includegraphics[width=14cm]{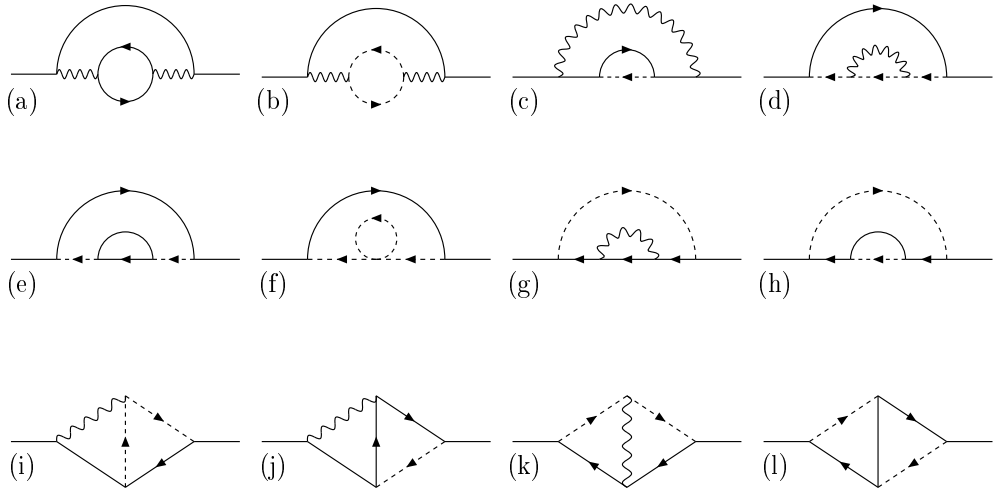}
\end{center}
\caption{ 
\small Two-loop $O(\alpha_s^2)$ contributions to the gluino self energy, 
with quark and squark propagators. Other diagrams obtained 
by charge conjugation are not shown. 
}
\label{fig2}
\end{figure}

The two-loop $O(\alpha_s^2)$ correction 
$\delta m_{\sg}^{(2)}$ consists of two parts, 
$\delta m_{\sg}^{(2)}=\delta m_{\sg}^{(2,1)}+\delta m_{\sg}^{(2,2)}$, 
where $\delta m_{\sg}^{(2,1)}$ is the contribution of the diagrams 
with only gluons and gluinos, while $\delta m_{\sg}^{(2,2)}$ is 
the remaining contribution with quark and squark loops. 
Two-loop self energy diagrams for these contributions 
are shown in Fig.~\ref{fig1} and Fig.~\ref{fig2}, respectively. 
In these figures, 
the wavy line, solid line without an arrow, solid line with an arrow, and 
dashed line with an arrow represent the gluon, gluino, quark, and squark, 
respectively. 

The contribution $\delta m_{\sg}^{(2,1)}$ 
is obtained by applying the formula of the $O(\alpha_s^2)$ QCD 
correction to the quark masses 
in the \dr scheme \cite{avdeev} to a fermion 
in SU(3) adjoint representation. The result is  
\beq
\delta m_{\sg}^{(2,1)}= 
\left( \frac{C_V \alpha_s}{4\pi} \right)^2 M_3 
\left( -48\log\frac{M_3}{Q} + 36\log^2\frac{M_3}{Q} +26+5\pi^2 -4\pi^2\log 2 
+6\zeta_3 \right) , 
\label{YMpart} 
\eeq
where $\zeta_3=\sum_{n=1}^{\infty}n^{-3}\simeq 1.202$. 
We have verified Eq. (\ref{YMpart}) by explicit calculation of the 
diagrams. At $Q=M_3$, the correction (\ref{YMpart}) is 
$\delta m_{\sq}^{(2,1)}/M_3\sim 31(\alpha_s/\pi)^2\sim 0.03$. 

The contribution $\delta m_{\sg}^{(2,2)}$ including quark and squark 
loops is calculated as follows: 
The Feynman integrals are decomposed into 
basic scalar integrals given in Refs.~\cite{tarasov,martin}, 
with the help of the integration by parts 
technique~\cite{tarasov,tarcer,grozin}. 
The resulting formulas are 
then numerically evaluated by the package TSIL \cite{tsil} for 
two-loop integrals. 
We have analytically checked that the $Q$ 
dependences of $\delta m_{\sg}^{(2,1)}$ in Eq.~(\ref{YMpart}) and 
$\delta m_{\sg}^{(2,2)}$ are consistent with the two-loop renormalization 
group equation \cite{mv1,rge} of $M_3$. 

The explicit form of $\delta m_{\sg}^{(2,2)}$ 
is rather long and will be presented elsewhere. 
Here we just show, for reference, the form in the 
limit of $m_{\sq}\gg M_3$, obtained by the 
heavy mass expansion technique \cite{HME}: 
\bea
\lefteqn{ \delta m_{\sg}^{(2,2)}(m_{\sq}\gg M_3) =} && 
\nonumber \\ 
&& 
\frac{\alpha_s^2M_3}{(4\pi)^2} 
\left[ 72 \log^2\frac{m_{\sq}}{Q} + 242 \log\frac{m_{\sq}}{Q} 
+\log\frac{M_3}{Q}\left( 54 - 288 \log\frac{m_{\sq}}{Q} \right) 
-172 + \frac{14}{3}\pi^2 \right] 
\nonumber\\ 
&& +\frac{\alpha_s^2M_3}{(4\pi)^2} N_q C_VT_F 
\left( -8\log^2\frac{M_3}{Q} + \frac{52}{3}\log\frac{M_3}{Q}
-\frac{37}{3}-\frac{4}{3}\pi^2  \right) . 
\label{asympt}
\eea
The last line of Eq.~(\ref{asympt}), which is independent of $m_{\sq}$, 
comes from the diagram (a) in Fig.~\ref{fig2}. 
We have checked that the $m_{\sq}$ dependence 
of Eq.~(\ref{asympt}) is consistent with 
the two-loop running of the gluino mass in the 
effective theory where squarks are integrated out \cite{split}. 

We then present numerical results of the mass correction to 
the gluino, for the running tree-level mass $M_3(M_3)=580$ GeV 
which is close to the values in the SPS1a point. 
The strong coupling constant 
is determined by $\alpha_s(m_Z)=0.12$, which is the running parameter 
within the standard model. 

\begin{figure}[ht]
\begin{center}
\includegraphics[width=10cm]{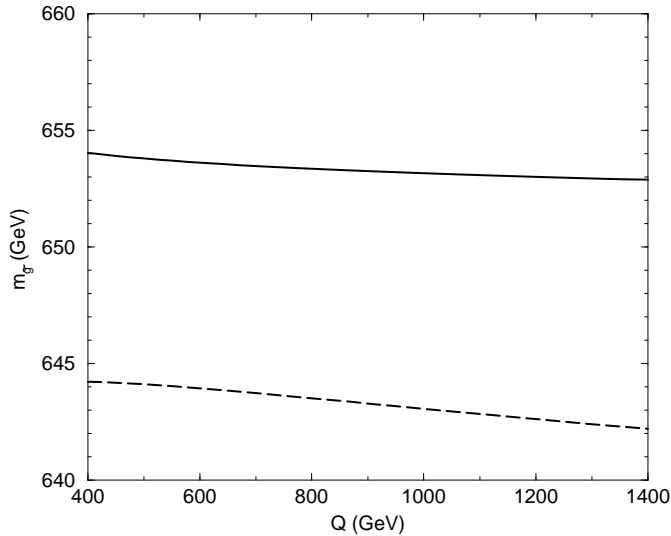}
\end{center}
\caption{ 
\small Dependence of the one-loop (dashed) and 
two-loop (solid) pole masses of the gluino 
on the renormalization scale $Q$. 
Mass parameters are $(M_3,m_{\sq})=(580,800)$ GeV at $Q=580$ GeV. 
}
\label{figqren}
\end{figure}

In Fig.~\ref{figqren}, we show the dependence of the 
one-loop pole mass $m_{\sg}^{(1)}=M_3+\delta m_{\sg}^{(1)}$ 
and two-loop pole mass 
$m_{\sg}^{(2)}=M_3+\delta m_{\sg}^{(1)}+\delta m_{\sg}^{(2)}$ 
on the renormalization scale $Q$. 
We give running parameters $M_3(Q_0)=580$ GeV and 
$m_{\sq}(Q_0)=800$ GeV at $Q_0=580$ GeV, and evolve them and $\alpha_s$ 
to a given $Q$ by $O(\alpha_s^2)$ renormalization group equations. 
For reference, the tree-level mass $M_3(Q)$ decreases from 
589 GeV at $Q=400$ GeV to 559 GeV at $Q=1400$ GeV.  
We see that the $Q$ dependence slightly improves as one includes 
$\delta m_{\sg}^{(2)}$. One should however note that  
$\delta m_{\sg}^{(2)}$ is much larger than the $Q$-dependence of the 
one-loop result $m_{\sg}^{(1)}$. It clearly shows that the 
latter is not an adequate estimate of the higher-order contribution 
to the pole mass. This property has already been observed for the 
$O(\alpha_s^2)$-correction to the squark mass \cite{msq2loop}. 

\begin{figure}[ht]
\begin{center}
\includegraphics[width=10cm]{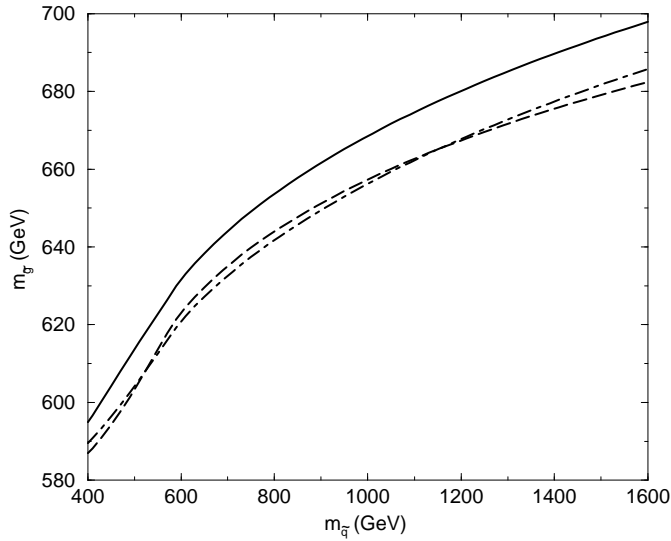}
\end{center}
\caption{ 
\small The pole masses of the gluino at the one-loop (dashed), 
two-loop (solid), and modified one-loop mass Eq.~(\ref{improve}) (dot-dashed), 
for the tree-level mass $M_3(M_3)=580$ GeV, as functions of $m_{\sq}(M_3)$. 
}
\label{fig4}
\end{figure}

In Fig.~\ref{fig4}, we compare $m_{\sg}^{(1)}$ and $m_{\sg}^{(2)}$ 
as functions of the running squark mass $m_{\sq}(Q=M_3)$, 
for $M_3(M_3)=580$ GeV. Here the renormalization scale is 
fixed at $Q=580$ GeV. 
The two-loop correction $\delta m_{\sg}^{(2)}$ is positive 
and in the range of $8-15$ GeV for $m_{\sq}=400-1600$ GeV. 
This correction is $O(1-2)$ \% of the one-loop result $m_{\sg}^{(1)}$ 
and somewhat larger than the expected uncertainty 
(8.0 GeV to 6.5 GeV) in the gluino mass determination 
at future colliders \cite{msgdet1}. 
For $m_{\sq}>M_3$, $\delta m_{\sg}^{(2)}$ increases with 
$m_{\sq}$, as suggested by the $m_{\sq}\gg M_3$ limit (\ref{asympt}). 

We also show the result of the modified one-loop 
formula of the pole mass 
\beq
m_{\sg}^{(1)}({\rm mod.}) = M_3(Q) + 
\frac{C_V \alpha_s(Q)}{4\pi} m_{\sg} \left( 5 -6\log \frac{m_{\sg}}{Q} \right) 
+\frac{\alpha_s(Q)}{\pi} N_q T_F m_{\sg} 
B_1(m_{\sg}^2, 0, m_{\sq}({\rm pole})) ,
\label{improve}
\eeq
where running masses in Eq.~(\ref{oneloop}) are replaced by 
the pole masses. Eq.~(\ref{improve}) includes 
higher-order corrections by one-loop 
renormalization group equations \cite{pierce2}. 
In Fig.~\ref{fig4}, it is seen that the modification (\ref{improve}) 
does not work for the inclusion of the two-loop 
correction $\delta m_{\sg}^{(2)}$. It would be a useful task to 
find other modifications of the one-loop formula which 
incorporate leading part of $\delta m_{\sg}^{(2)}$. 

We finally comment on the effects of the 
quark masses $m_q$ and left-right mixings of squarks, 
both of which are ignored here, to the gluino mass correction. 
Since these parameters break the SU(2)$\times$U(1)
gauge symmetry, their contributions to $m_{\sg}$ 
should be suppressed by factors $m_q^2/m_{\sq}^2$ or $m_q^2/m_{\sg}^2$ 
compared to the SU(2)$\times$U(1)-symmetric contribution shown in 
this paper. 
We therefore expect that, when the gluino and squarks are 
sufficiently heavy, their effects on 
the $O(\alpha_s^2)$ correction $\delta m_{\sg}^{(2)}$ 
are numerically irrelevant 
for future precision studies of the SUSY particles. 
However, these effects might become relevant for 
SUSY parameter sets with relatively light gluino or squarks. 
We will present the complete result of the 
$O(\alpha_s^2)$ mass correction  including 
these effects, as well as the 
$O(\alpha_s h_q^2)$ contributions involving 
quark-Higgs Yukawa couplings $h_q$, elsewhere. 

In conclusion, we have calculated the two-loop SUSY QCD contribution 
to the gluino pole mass by diagrammatic method, for the gluino 
and squarks sufficiently heavier than the quarks. 
The $O(\alpha_s^2)$ correction to the gluino mass has 
been shown to be typically $1-2$ \%. 
For the case of $M_3(M_3)=580$ GeV, 
this correction is similar to, or larger than, 
the expected uncertainty in the mass determination from 
precision measurements at future colliders. 
This correction would affect the extraction of $M_3$ from 
experimental data and, since $M_3$ contributes 
to the running of many soft SUSY breaking parameters, 
the determination of the SUSY breaking at the unification scale. 

{\it Acknowledgements:} 
The author thanks Mihoko Nojiri for discussion from 
which this work has originated, 
useful suggestions and comments, 
and careful reading of the manuscript.


\end{document}